\documentclass[twocolumn]{aa}

  \usepackage{amscd}
  \usepackage{amsmath}
  \usepackage{amssymb}
  \usepackage{aas_macros}
  \usepackage[T1]{fontenc}
  \usepackage[varg]{txfonts}
  \usepackage{graphicx}
  \usepackage{natbib}
  \usepackage{verbatim}
  \usepackage{array}
\begin{document}

\title{The case against the progenitor's carbon-to-oxygen ratio as a
  source of peak luminosity variations in Type Ia supernovae}
\titlerunning{C/O ratio of the progenitor and SN Ia luminosity}

\author{F. K. R{\"o}pke\inst{1},
        \and
        W. Hillebrandt\inst{2}}
          
   \offprints{F. K. R{\"o}pke}

   \institute{Max-Planck-Institut f\"ur Astrophysik,
              Karl-Schwarzschild-Str. 1, D-85741 Garching, Germany\\
              \email{fritz@mpa-garching.mpg.de}
              \and
              \email{wfh@mpa-garching.mpg.de}
             }

\abstract{One of the major challenges for theoretical modeling of Type
  Ia supernova explosions is to explain the diversity of these events
  and the empirically established correlation between their peak
  luminosity and light curve shape. In the framework of the
  so-called Chandrasekhar mass models, the progenitors
  carbon-to-oxygen ratio has been suggested to be a principal source
  of peak luminosity variations due to a variation in
  the production of radioactive $^{56}$Ni during the explosion. The
  underlying idea is that an
  enhanced carbon mass fraction should result in a more vigorous
  explosion since here the energy release from nuclear reactions is
  increased. It was suspected that this would produce a higher
  amount of  $^{56}$Ni in the ejecta. In this letter we describe a
  mechanism resulting from an interplay between nucleosynthesis and
  turbulent flame evolution which counteracts such an effect. Based on
  three-dimensional simulations we argue that it is nearly balanced
  and only minor differences in the amount of
  synthesized $^{56}$Ni with varying carbon mass fraction in the
  progenitor can be expected. Therefore this progenitor parameter
  is unlikely to account for the observed variations in Type Ia
  supernova luminosity. We discuss possible effects on the calibration
  of cosmological measurements.}

\maketitle


\section{Introduction}
\label{intro_sect}

The question of the origin of the observed diversity of Type Ia
supernovae (SNe Ia) has been a long standing problem. Although these
astrophysical events possess a surprising homogeneity of features,
which led to the application of SNe Ia as distance indicators in
cosmology, variations of  photometric and
spectroscopic data have been reported. Nevertheless,
corrections of the luminosity based on an empirical
relation between peak luminosity and light curve shape
\citep{phillips1993a} made SNe Ia a
major tool of observational cosmology (for a recent review see
\citealt{leibundgut2000a}).
A theoretical understanding of this correlation is still lacking. Thus
the challenge for astrophysical theory is to construct a SN Ia
explosion model that is robust and---as a first step---able to explain
the observed variations.

It is generally agreed on the fact that SNe Ia originate from
thermonuclear explosions of white dwarf (WD) stars, although many
scenarios have been proposed for the particular realization (see
\citet{hillebrandt2000a} for a recent review).
The currently favored model is a binary system in which a WD composed
of carbon and oxygen
accretes matter from a non-degenerate companion until it reaches the
Chandrasekhar mass. At this point, thermonuclear burning at the center
of the star develops into a flame which propagates outward. From the
viewpoint of hydrodynamics, two modes of flame propagation are
possible: a subsonic deflagration, in which the reaction is mediated
by thermal conduction of the degenerate electrons, and a supersonic
shock-induced detonation. \cite{arnett1969a} noted that to produce the
observed intermediate mass elements a prompt detonation can be
excluded. Therefore the flame starts out in the deflagration mode and
may or may not develop into a detonation. The key feature of this
model is that buoyancy-induced instabilities lead to a turbulization
of the flame front in the deflagration mode. Only due to this effect
the flame is accelerated 
sufficiently to power SN Ia explosions.

\begin{figure*}[t]
\centerline{
\includegraphics[width = 0.73 \linewidth]
  {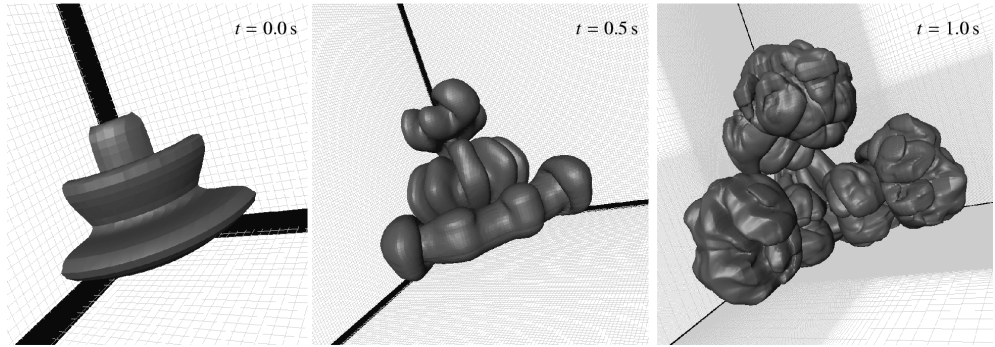}}
\caption{Temporal evolution of the burning front.
  \label{evo_fig}}
\end{figure*}

In the first stages the prevailing densities and temperatures are high
enough that the reaction terminates in nuclear statistical equilibrium
(NSE), consisting mainly of iron group elements. As the WD expands,
density and temperature drop and the thermonuclear reactions
produce intermediate mass elements such as Si, S, and
Ca.

The light curve of SNe Ia is powered by the radioactive decay of $^{56}$Ni and
$^{56}$Co. The peak luminosity is a measure of the $^{56}$Ni produced by
the explosion (``Arnett's law'', \citealt{arnett1982a}).
Parameters that have frequently been suggested to affect the amount of
$^{56}$Ni synthesized are the carbon-to-oxygen (C/O) ratio of the WD,
its central density prior to ignition, and its metallicity. Of course,
other parameters 
like rotation and asphericity of the explosion may play a role, too.
A thorough investigation of the impact of these parameters is
mandatory for the validation of cosmological measurements, since here
possible evolution effects with cosmic age are critical.

The objective of the present study is to explore the effect of the
progenitor's C/O ratio on the supernova explosion by means of
three-dimensional hydrodynamical models. Since we are mainly
interested in the explosion energy and the amount of synthesized
$^{56}$Ni, it is justified to focus on the
deflagration stage, leaving aside a possible delayed detonation. Only
in this first stage of the SN Ia explosion the prevailing densities and
temperatures are sufficient for burning to iron group elements, and therefore the
main part of the $^{56}$Ni and probably also of the explosion energy is
produced here. 

\section{The numerical model}

The numerical model applied to simulate the thermonuclear explosion is
the same as used for several studies by
\citet{reinecke2002b,reinecke2002d}. Therefore we will
be short in the description of the model and only mention the basic
aspects here. 

The vast range of involved length scales makes fully resolved SN Ia
explosion simulations impossible. 
Therefore we describe
the flame propagation by
applying the level set method \citep{osher1988a}. Neither the
internal flame structure nor its wrinkling on small scales are
resolved but the flame is rather modeled as a discontinuity
separating fuel and ashes. This discontinuity represents the mean
position of the flame. To track its propagation it is associated with
the zero level set of a scalar field $G$ which is evolved according to the
scheme described by \citet{reinecke1999a}. In this scheme the
effective flame velocity must be provided and for this we employ the
subgrid scale model proposed by \citet{niemeyer1995b} which describes
the effects of the turbulent motions on unresolved scales.
The hydrodynamics is modeled based on the PROMETHEUS implementation
\citep{fryxell1989a} of the piecewise parabolic method
\citep{colella1984a}. 

Due to the restricted computational resources only a very simplified
description of the nucleosynthesis is possible concurrent with the
explosion simulation. We follow the approach suggested by
\citet{reinecke2002b}, who include five species, namely
$\alpha$-particles, $^{12}$C, $^{16}$O, $^{24}$Mg as a representative
of intermediate mass elements and $^{56}$Ni as a representative of iron
group nuclei (denoted as ``Ni'' in the following to avoid confusion
with the particular isotope as part of the ejecta). The fuel is assumed to consist of a mixture of carbon and
oxygen. At the initially high densities burning proceeds to
NSE composed of $\alpha$-particles and
``Ni''. Depending on temperature and density in the ashes, the
amount of $\alpha$-particles and nickel changes. Once the
fuel density drops below
$5.25 \times 10^7 \,\mathrm{g}\,\mathrm{cm}^{-3}$ due to the expansion
of the WD, burning is assumed
to terminate at intermediate mass elements and below $1 \times 10^7
\,\mathrm{g}\,\mathrm{cm}^{-3}$ burning becomes very slow and is not
followed anymore.

In this letter we will present three simulations that span one octant of
our coordinate system and assume mirror symmetry to the other
octants. The simulations were set up on a cartesian computational
grid that was equally spaced in the inner regions. In order to be able
to follow the explosion for a longer period of time in the outer parts
the width of the grid cells was expanded exponentially.

The resolution of the runs was rather low---the
computational domain was divided in $[256]^3$ grid cells corresponding
to a central grid resolution of $10^6 \, \mathrm{cm}$. In each
direction the grid length in the outer 35 zones was increased
subsequently by a factor of 1.15. The burning was centrally ignited
and the spherical initial flame structure was perturbed with
three toroidal rings per octant (see also Fig.~\ref{evo_fig}).

\section{Results of three-dimensional explosion models}

Snapshots from the explosion simulation for
$X(^{12}\mathrm{C}) = 0.46$ are given in
Fig.~\ref{evo_fig}. Here the position of
the flame front is rendered as represented by the zero level set of
the scalar field $G$ dividing fuel and ash. The spatial extent of the burnt region
can be inferred from the plotted grid which also visualizes our setup with
uniform grid cells in the inner region and an exponential growth of
the grid spacing further out. The flame as initialized in our setup is
shown in the left snapshot of Fig.~\ref{evo_fig}. In the
subsequent images the growth of instabilities and an increasing
wrinkling of the flame front is visible.

\begin{figure}[ht]
\centerline{
\includegraphics[width = 0.97\linewidth]
  {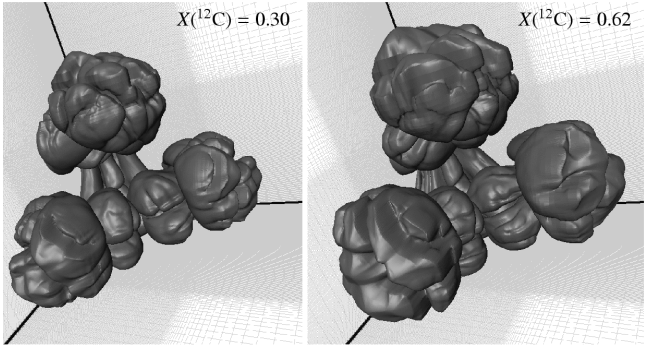}}
\caption{Flame front at $t=1.0 \, \mathrm{s}$ for models with
  different carbon mass fraction of the progenitor material.
  \label{co_fig}}
\end{figure}

\begin{table}
\centering
\caption{Characteristics of the explosion models.
\label{results_tab}}
\setlength{\extrarowheight}{2pt}
\begin{tabular}{llll}
\hline\hline
$X(^{12}\mathrm{C})$ & $E_\mathrm{nuc}$ $[10^{50}\,\mathrm{erg}]$ &
$M$\/(``Ni'') $[M_\odot]$ &
$\max(M_\alpha)$ $[M_\odot]$\\
\hline
0.30 & 8.85 & 0.5178 & 0.04579 \\
0.46 & 9.46 & 0.5165 & 0.05177 \\
0.62 & 9.97 & 0.5104 & 0.05636 \\
\hline
\end{tabular}
\end{table}

To test the effect of a varying carbon mass fraction of the WD, we
performed three simulations with values of $X(^{12}\mathrm{C})$ given
in Table~\ref{results_tab}.
Figure~\ref{co_fig} shows snapshots from models with different
$X(^{12}\mathrm{C})$ after $1.0 \, \mathrm{s}$. It is essential for
the following argumentation to note that for all three models the
extent of the burnt region and also the flame morphology are
surprisingly similar. The latter is in agreement with the
findings of \citet{gamezo2003a}, but \cite{khokhlov2000a} claims that a
decreasing $X(^{12}\mathrm{C})$ would result in weaker explosions
because of a delay of the development of the
Rayleigh-Taylor instability, which seems to be only a minor effect in
our simulations.

The energy release of the different models are summarized
in Table~\ref{results_tab}. Obviously, a higher carbon
mass fraction leads to a higher total energy production for fixed
central densities. 
This trend is not surprising. For
larger $X(^{12}\mathrm{C})$ the laminar burning velocity increases
\citep{timmes1992a}. This effect, however,
will be important only in the earliest stages of the explosion,
since flame propagation becomes determined by turbulence shortly after
ignition and can therefore only account for minor changes.
More significant, a higher carbon mass fraction will increase the total energy
generation for the simple reason that the binding energy of
$^{12}\mathrm{C}$ is lower than that of $^{16}\mathrm{O}$ so that it
releases more energy by fusion to ``Ni''.

Therefore the very similar flame evolution of all models is not expected in
a simple picture. That this is nevertheless the case is corroborated
by the following arguments.
As mentioned in Sect.~\ref{intro_sect}, the flame evolution in the
deflagration mode is dominated by turbulization which results from
Rayleigh-Taylor-like instabilities.
An important parameter for this is the effective gravity
experienced by the flame. It
accelerates the development of flame structures resulting from the
non-linear stage 
of the Rayleigh-Taylor instability.
At early stages the energy fed into the turbulent cascade
from the interfaces of large scale
buoyant burning bubbles will be enhanced by a higher $g$.
Thus, the turbulent
cascade will develop more quickly and the turbulence-induced boost of
the effective flame propagation velocity sets in earlier.

\begin{figure}[ht]
\centerline{
\includegraphics[width = 0.71 \linewidth]
  {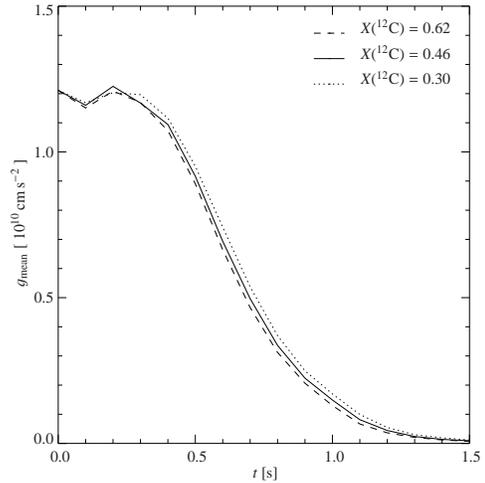}}
\caption{Mean gravitational acceleration $g_\mathrm{mean}$ experienced
  by the flame front for models with different carbon mass fraction of
  the progenitor material. 
  \label{geff_fig}}
\end{figure}

Our models, however, do not show such a behavior.
In Fig.~\ref{geff_fig} the temporal evolution of the mean $g$
experienced by the flame is plotted. We note only small differences
for varying $X(^{12}\mathrm{C})$. This results in a very similar
evolution of the turbulent energy in our models.

\begin{figure}[ht]
\centerline{
\includegraphics[width = 0.97 \linewidth]
  {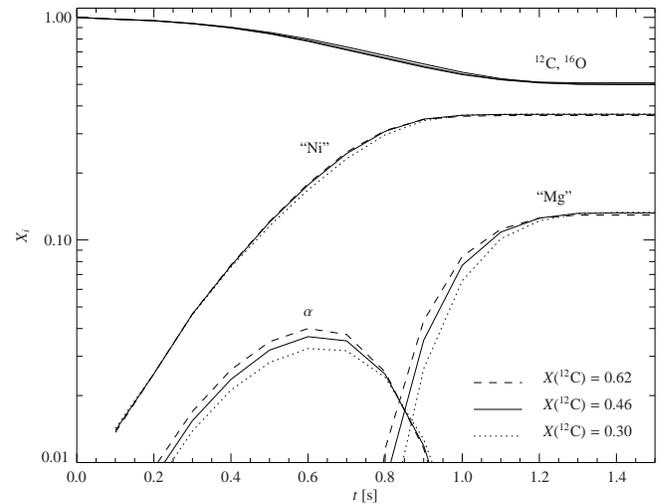}}
\caption{Temporal evolution of the chemical composition in  models with
  different carbon mass fraction of the progenitor material.
  \label{evalmass_fig}}
\end{figure}

The mechanism that is responsible for the reduced effect of the
varying carbon mass fraction on the
flame propagation becomes plausible from
the evolution of the chemical composition of our models (see
Fig.~\ref{evalmass_fig}). Though the temporal evolution of
the mass fractions of carbon and oxygen as well as nickel show little
differences, the variations of the mass fractions
of $\alpha$-particles and intermediate mass elements is significant. 

As pointed out by \citet{reinecke2002b}, the abundance of
$\alpha$-particles
is important for the explosion dynamics. At high densities and
temperatures the flame converts the C/O fuel to NSE
which in our models is represented by a mixture of nickel (representing
the iron group nuclei) and $\alpha$-particles. 
The $\alpha$-particles are produced at high temperatures in the ashes and
have two effects. First, the binding energy of the ashes is lower in
case of higher $X(\alpha)$, so less energy is released. Second,
the number of particles per unit mass of the ashes increases and this
decreases the temperature. Both effects result in  a lower temperature
and a higher density of the burning products which delays the expansion of the
WD and decreases the buoyancy of the
burnt regions. Thus the flame acceleration is  lower and the burning
is suppressed. With further expansion of the WD, the
$\alpha$-particles are converted to nickel. 

In our simulations
$\alpha$-particles are present between $0.2 \, \mathrm{s}$ and
$0.9\,\mathrm{s}$. The maximal $X(\alpha)$ is given in
Table~\ref{results_tab}.
Obviously,  higher
carbon mass fraction in the fuel gives rise to a larger fraction of
$\alpha$-particles in the ashes. Therefore the evolution of the
flame front is more delayed and consequently the explosion
dynamics of the models in the first stage when the iron group elements
are synthesized is comparable. The result of this effect is a
production of similar amounts
of iron group elements. The energy stored in the $\alpha$-particles
is released at later times and the total energy released in the
explosion models varies for about 12\% while the mass of produced
nickel differs only about 1.4\% (cf.\ Table~\ref{results_tab}).

Note that the explosion energies and the masses of synthesized iron
group elements in all our models are rather on the low side to
explain a prototype SNe Ia. This is due to the low numerical
resolution of our models. Although they are expected to be numerically
converged \citep{reinecke2002c}, the resolution does not allow to apply multi-spot
ignition scenarios which have been shown to produce more vigorous
explosions \citep{reinecke2002d}. However, the trends inferred from our models
are expected to be robust.

\section{Conclusions}

In this letter we described a mechanism that is responsible for the
somewhat surprising fact that although the progenitor's C/O ratio
affects the energy release of SN Ia explosions, it has little effect
on the peak luminosity determined by the $^{56}$Ni mass. This is found
with help of
three-dimensional simulations and disagrees with the results
reported from one-dimensional models. \citet{hoeflich1998a} find
 a 14\% decrease in $M_\mathrm{Ni}$ when reducing the C/O ratio from
 1/1 to 2/3; the fact that they considered a delayed detonation model
 should not greatly affect the comparability for reasons given in
 Sect.~\ref{intro_sect}. The
``working hypothesis'' of an increased $^{56}$Ni production with higher
$X(^{12}\mathrm{C})$ established by \citet{umeda1999b} cannot be
confirmed. The reason for this discrepancy is a complex interplay between
nucleosynthesis and nonlinear flame evolution which can be modeled
appropriately only in three dimensions.
The models presented here are based on a rather simplistic description
of the nuclear reactions, which will be improved in forthcoming
models. Nevertheless, the features that are important for the
explosion dynamics are taken into account and trends can be revealed
with this approach.

An interesting question is how our results affect the peak
luminosity--light curve shape relation of SNe Ia. Since no light
curves were calculated from our models we can only speculate on the
effects based on the trends found by \citet{arnett1982a} and
\citet{pinto2000a} by means  of analytic studies. According to
``Arnett's law'' the peak luminosity
should reflect the amount of radioactive $^{56}$Ni synthesized in the
explosion. Since in our models the evolution of the densities of the
ashes are very similar, the little variation in the iron group elements
will result in little variation of the $^{56}$Ni mass. Thus the peak
luminosity should be roughly the same for our three models. The
explosion energy, however, varies significantly and a greater
expansion at late stages of the explosion associated with a larger C/O
ratio would result in a more rapid
decline of the light curve but also in a somewhat higher
luminosity. This is in contradiction to the empirically established
relation that for brighter SNe Ia the light curve declines slower,
which has been extensively applied to calibrate cosmological distance
measurements.

\citet{timmes2003a} analytically
found a linear dependence of the amount of produced $^{56}$Ni on the
progenitor's metallicity. This in turn is, however, not likely to
affect the explosion energy and dynamics significantly. Of course,
from the viewpoint of stellar evolution
there is a interrelation between the progenitor's metallicity and
other parameters, including its carbon mass fraction. 
Therefore it is plausible that the ``luminosity-width relation'' is caused
by a combination of different parameters and that a direct relation
between the $^{56}$Ni mass and the peak luminosity might be
oversimplified. It seems well possible that the distribution of the
$^{56}$Ni due to three-dimensional effects plays an important role here.
In a subsequent publication
we will report on a systematic parameter study addressing these
questions.
Ultimately, SN Ia explosion models will have to be coupled to
realistic stellar evolution of the progenitor system and the
calculation of synthetic light curves becomes mandatory.


\begin{acknowledgements}
We thank M.~Reinecke, C.~Travaglio, M.~Gieseler, and W.~Schmidt for
helpful discussions.
\end{acknowledgements}

\end{document}